# Efficient compressive sensing for machinery vibration signals

Imen Tounsi [1,2], Fadi Karkafi [1], Mohammed El Badaoui [1,2], François Guillet [2]

1 Safran Tech, Rue des Jeunes Bois – Châteaufort 78772 Magny–les–Hameaux, France

2 UJM-St-Etienne, LASPI, UR3059, F-42023, Saint-Etienne, France

**Abstract.** Mechanical vibration monitoring often requires high sampling rates and generates large data volumes, posing challenges for storage, transmission, and power efficiency. Compressive Sensing (CS) offers a promising approach to overcome these constraints by exploiting signal sparsity to enable sub-Nyquist acquisition and efficient reconstruction. This study presents a comprehensive comparative analysis of the key components of the CS framework: sparse basis, measurement matrix, and reconstruction algorithm for machinery vibration signals. In addition, a hardware-efficient measurement matrix, the Wang matrix, originally developed for image compression, is introduced and evaluated for the first time in this context. Experimental assessment using the HUMS2023 and the CETIM gearbox datasets demonstrates that this matrix achieves superior reconstruction quality, with higher SNR, compared to conventional Gaussian and Bernoulli matrices, especially at high compression ratios.

## 1  Introduction

Mechanical components play a vital role in propulsion and actuation systems, ensuring efficient power transmission, load management, and motion control [1]. Their reliability is critical, as failures can compromise safety, increase maintenance costs, and disrupt missions [2]. Vibration analysis is a widely used technique for health monitoring [3], but extracting fault features remains challenging due to high-frequency noise and large data volumes from high sampling rates and multiple sensors [4]. In many practical environments, such as aerospace applications, strict limitations on bandwidth, storage, and power further highlight the need for efficient data compression to ensure reliable and resource-efficient condition monitoring.

Compressive sensing (CS) offers an effective solution to these challenges. By exploiting signal sparsity in appropriate transform domains, CS enables sub-Nyquist sampling while preserving key fault features [5]. This reduces data redundancy and resource demands, making it well-suited for real-time applications. Research on CS for rotating machinery has mainly focused on three components: the sparse basis, measurement matrix, and reconstruction algorithm. Early studies used analytical bases such as the Discrete Cosine Transform (DCT), Discrete Fourier Transform (DFT), and wavelets for their simplicity and energy compaction (e.g., Wang et al., 2015 [6]; Bai et al., 2022 [7]). To enhance adaptability, Chen et al. (2014) [8] introduced K–Singular Value Decomposition (K-SVD) dictionary learning, and Kida and Shinozaki (2019) [9] proposed the Order-Ratio Basis (ORB) to capture rotational periodicity. Measurement matrices are often Gaussian or Bernoulli due to their theoretical guarantees under the Restricted Isometry Property (RIP) (Wu et al., 2019, 2021 [10,11]), while Wang and Xia (2024) [2] developed a noise-robust random convolution matrix using a generative flow model. For reconstruction, greedy algorithms such as Orthogonal Matching Pursuit (OMP) and Compressive Sampling Matching Pursuit (CoSaMP) are popular for their efficiency (Bai et al., 2022; Wu et al., 2019 [7,10]), whereas recent generative and learning-based models, such as GLOW (Wang & Xia, 2024 [4]), achieve higher fidelity at the cost of greater computational complexity. Overall, prior studies demonstrate the promise of CS for mechanical vibration analysis but also expose key limitations. Most works assess only a single combination of sparse basis, measurement matrix, and reconstruction algorithm, lacking systematic evaluation of their joint influence on performance. Moreover, many recent methods rely on dataset-specific or learned representations that may not generalize across machines or operating conditions. Although Gaussian and Bernoulli matrices offer strong theoretical guarantees, their computational demands limit real-time or embedded aerospace applications requiring hardware simplicity and energy efficiency [12]. Gaussian matrices employ randomly distributed real-valued elements, requiring numerous multiplications and summations during measurement acquisition. This increases computational complexity and power consumption and risks exceeding the dynamic range of the analog-to-digital converter (ADC) [12]. Bernoulli matrices, composed solely of +1 and -1 entries, eliminate multiplications but still involve extensive summations that can produce large measurement magnitudes.

To address these limitations, a novel measurement matrix, the Wang matrix [12], originally proposed for image applications and optimized for hardware implementation, is introduced for the first time in this context. The results demonstrate that this matrix not only enhances hardware efficiency but also achieves superior reconstruction quality compared to conventional Gaussian and Bernoulli random measurement matrices. The proposed method is further evaluated using two real-world datasets from HUMS2023 and CETIM gearbox datasets, demonstrating its robustness across different operating conditions.

## 2    Compressive sensing framework

CS provides a powerful framework for efficient signal acquisition and reconstruction. Unlike traditional Nyquist-based sampling, CS exploits the inherent sparsity of signals in an appropriate transform domain to recover them accurately from a smaller number of measurements. This recovery is achieved through nonlinear optimization techniques. Fig. 1 provides a schematic representation of the CS workflow showing three key components: sparse representation $s$, compressed sampling $y$, and signal reconstruction $\hat{x}$.

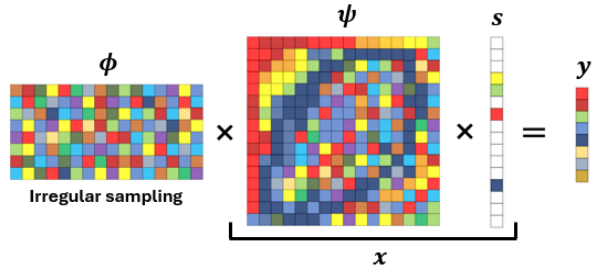

**Fig. 1 Schematic diagram of the compressive sensing algorithm**

### 2.1 Sparsity and incoherence in compressive sensing

Sparsity refers to the property of a signal whose representation in an appropriate basis contains only a few informative coefficients that stand out significantly from the background noise, while the remaining coefficients are negligible or close to zero [5]. This means that the signal can be expressed with only $K$ non-zero coefficients, where $K \ll N$, and $N$ denotes the dimensionality of the original signal. In a chosen basis $\boldsymbol{\psi} \in \mathbb{R}^{N \times N}$, the signal is said to be $K$-sparse and can be represented as:

$$x = \boldsymbol{\psi} s \quad (1)$$

$s$ denotes the sparse representation of the original signal in the transform domain. CS reduces sampling requirements by capturing a signal's key information via a designed measurement matrix $\boldsymbol{\phi} \in \mathbb{R}^{M \times N}$, where M is the number of compressed measurements [5]. Then, the compressed measurement vector can be expressed as follows:

$$y = \boldsymbol{\phi} x = \boldsymbol{\phi} \boldsymbol{\psi} s \quad (2)$$

To ensure efficient CS, the Wang matrix must exhibit low coherence with the sparsity basis. The coherence [13] between these two matrices is given by:

$$\mu(\boldsymbol{\phi}, \boldsymbol{\psi}) = \sqrt{N} \max_{j,k} |\langle \phi_j, \Psi_k \rangle| \quad (3)$$

where $\phi_j$ is the j-th row of $\boldsymbol{\phi}$ and $\psi_k$ is the k-th column of $\boldsymbol{\psi}$.

The compression ratio (CR) represents the percentage of measurements retained after compression and is defined as:

$$CR = \frac{M}{N} \times 100 \quad (4)$$

### 2.2 Proposed methodology

The Wang matrix, introduced in [12], is a hardware-efficient measurement matrix specifically designed to simplify the image acquisition process in CS. Unlike Gaussian and Bernoulli matrices, which require extensive multiplications and summations, the Wang matrix is constructed through a random subsampling of the identity matrix. In other words, a subset of the rows of the identity matrix is randomly selected without replacement, as shown in (5), ensuring that each measurement corresponds to a unique element of the original signal. This structure means that each measurement directly captures one element of the input without any linear combination or additional computation. Consequently, it offers a simple and low-power implementation that is particularly suitable for hardware-based CS systems.

$$\Phi = \begin{bmatrix} 1 & 0 & 0 & \cdots & 0 & 0 \\ 0 & 1 & 0 & \cdots & 0 & 0 \\ 0 & 0 & 1 & \cdots & 0 & 0 \\ \vdots & \vdots & \vdots & \ddots & \vdots & \vdots \\ 0 & 0 & 0 & \cdots & 1 & 0 \\ 0 & 0 & 0 & \cdots & 0 & 1 \end{bmatrix}_{N \times N} \xrightarrow{\text{Random } M \text{ rows}} \begin{bmatrix} 0 & 0 & 0 & \cdots & 1 & 0 \\ 1 & 0 & 0 & \cdots & 0 & 0 \\ 0 & 1 & 0 & \cdots & 0 & 0 \\ \vdots & \vdots & \vdots & \ddots & \vdots & \vdots \\ 0 & 0 & 1 & \cdots & 0 & 0 \\ 0 & 0 & 0 & \cdots & 0 & 1 \end{bmatrix}_{M \times N} \quad (5)$$

Then, the CS reconstruction process using the Wang matrix follows the general optimization problem defined in (6) [5].

$$\min_{s} \|s\|_1 \quad s.t \quad \boldsymbol{\phi}\boldsymbol{\psi} s = y \quad (6)$$

The recovered signal is then obtained as: $\hat{x} = \boldsymbol{\psi} s$. This work employs the Orthogonal Matching Pursuit (OMP) algorithm due to its balance between accuracy and computational cost, making it suitable for real-time vibration signal analysis. The performance of the CS algorithm is then evaluated using the Signal-to-Noise Ratio (SNR), defined as follows:

$$SNR = 10 \log_{10} \left( \frac{\sum_{i=1}^{N} x_i^2}{\sum_{i=1}^{N} (x_i - \hat{x}_i)^2} \right) \quad (7)$$

Where $x_i$ is the i-th sample point of the original signal $x$, and $\hat{x}_i$ is the i-th sample point of the reconstructed signal $\hat{x}$.

## 3 Experimental evaluation and results

### 3.1 Case Study 1: HUMS2023 Gearbox Dataset

#### 3.1.1 Data description

The vibration data used in this study come from the HUMS2023 data challenge [14] by the defence science and technology group, Australia. The dataset was recorded from the main rotor gearbox of a Bell Kiowa 206B-1 (OH-58) helicopter. The gearbox has two reduction stages (a spiral pinion/bevel gear stage and a planetary stage) with an input speed of 6000 RPM and an output speed of 344 RPM. Vibration signals were collected from four accelerometers mounted on the gearbox housing and a once-per-revolution tachometer. The dataset includes 526 four-channel recordings representing the last seven days of testing under 125% torque (379 Nm). The input pinion speed was 100 Hz, and the output shaft speed was 5.73 Hz. Key gear-mesh frequencies include 1900 Hz for the input pinion/bevel gear and 568 Hz for the planetary-stage sun, planet, and ring gears.

### 3.1.2 Sparsity and incoherence analysis

As a first step, the sparsity of the vibration signals was evaluated in different transform domains: DCT, DFT, and Daubechies wavelets (Db2 and Db8). The sparsity was quantified as the percentage of coefficients whose absolute values are lower than 1% of the maximum coefficient magnitude in the corresponding transformed signal. Fig. 2 illustrates the sparsity values obtained for each domain using the data from day 22. The DCT and DFT domains exhibit the highest sparsity ($\approx 0.99$), while the wavelet domains are less sparse, with Db8 performing better than Db2. These results confirm that the DCT and DFT are the most suitable candidates for CS of gearbox vibrations, as they enable efficient energy compaction.

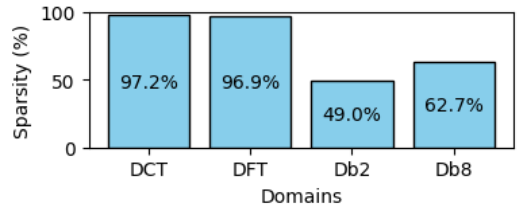

**Fig. 2 Sparsity values of the vibration signal from day 22 in different transform domains (DCT, DFT, Db2, Db8).**

Following the sparsity analysis, the incoherence between the measurement matrix and the sparse basis was evaluated to assess their suitability for CS. Table 1 summarizes the coherence values obtained for Gaussian, Bernoulli, and Wang measurement matrices in combination with the DCT and DFT domains.

**Table 1. Coherence values between different measurement matrices and sparse basis matrices.**

| Parameters | DCT | DFT |
|---|---|---|
| Gaussian | 5.57 | 4.19 |
| Bernoulli | 5.52 | 4.15 |
| Wang | 1.41 | **1.00** |

As shown in the table, the Gaussian and Bernoulli matrices exhibit relatively high coherence with the DFT and DCT bases. In contrast, the Wang matrix achieves the lowest coherence values ($\approx$ 1.0-1.4) with these transform domains, indicating strong mutual independence. A comparative performance analysis is conducted in the following section to further validate and confirm these preliminary observations.

### 3.1.3 Comparative evaluation of measurement matrices

To assess the impact of the measurement matrix on reconstruction quality, CS experiments were performed using DFT and DCT bases with the OMP algorithm. The three measurement matrices were compared in Fig. 3 by showing the variation of SNR between the original and reconstructed signals from day 22 across different CRs.

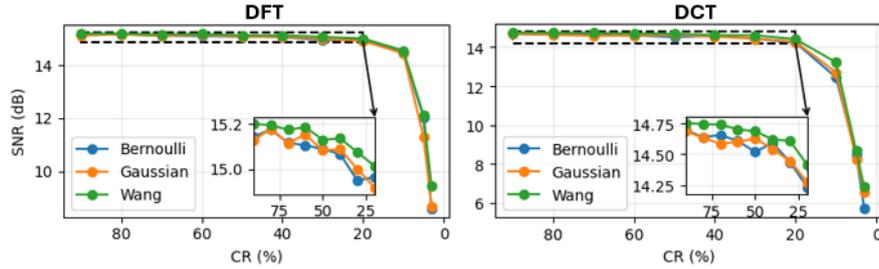

**Fig. 3 SNR vs. CR for a signal from day 22 using OMP in DFT (left) and DCT (right) domains with three measurement matrices (Bernoulli, Gaussian, Wang)**

Among the tested matrices, the Wang matrix consistently achieved superior reconstruction performance. It provided higher SNR values across all CRs compared to the other matrices. This indicates that it offers a more effective sampling structure for preserving signal information during the compression process.

Fig. 4 compares the original and reconstructed signals from day 22 (left), obtained using the DFT basis, Wang matrix, and OMP algorithm at a CR of 3%, demonstrating the high reconstruction accuracy of the CS approach. It also compares the effectiveness of the RMS health indicator on the original and reconstructed signal sets from day 22. As shown, the RMS values preserve the same overall trend, and the two peaks indicating the structure changes are clearly retained in the reconstructed signals.

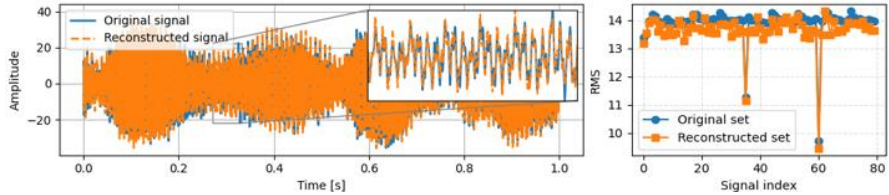

**Fig. 4 Comparison of original and reconstructed signals from day 22 (left) and RMS values of each signal from day 22 (right), with the original (blue) and reconstructed (orange) signals.**

An overall average SNR of 8.29 dB $\pm$ 0.31 dB, computed across all channels and days, underscores the robustness and reliability of the proposed reconstruction approach under a very low CR.

### 3.2 Case Study 2: CETIM Gearbox Dataset

### 3.2.1 Data description

The second set of experiments relies on the CETIM vibration dataset, which provides measurements acquired from a gearbox subjected to a gradual degradation process. The dataset includes twelve vibration signals, each corresponding to a distinct day of acquisition, thus capturing the evolution from a healthy to a faulty operating condition. The signals were sampled at 20 kHz over a duration of 3 seconds per acquisition. The dominant component in the vibration spectra is the gear meshing frequency, which varies slightly between 330 Hz and 346 Hz depending on the rotational speed.

### 3.2.2 Sparsity analysis

The sparsity of the CETIM vibration signals was evaluated in the same transform domains: DCT, DFT, Db2 and Db8. Fig. 5 presents the sparsity results for the signals acquired on day 1 (healthy) and day 12 (faulty). The DCT and DFT domains exhibit the highest sparsity in both conditions, confirming their strong ability to compact signal energy into a few dominant coefficients. However, when the defect develops (day 12), sparsity in these domains slightly decreases due to the appearance of impulsive and broadband components associated with the fault.

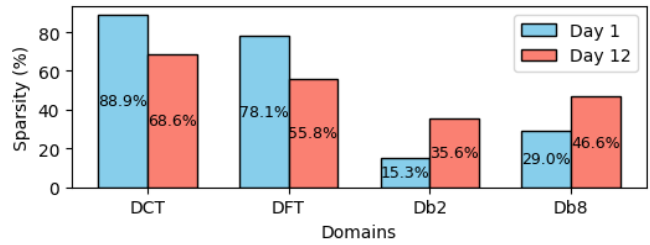

**Fig. 5** Sparsity values of the vibration signal from day 1 (in blue) and day 12 (in red) in different transform domains (DCT, DFT, Db2, Db8).

### 3.2.3 Comparative evaluation of measurement matrices

To assess the effect of the measurement matrix on reconstruction quality, CS experiments were performed on the CETIM data using DFT and DCT bases with the OMP algorithm. The three measurement matrices were compared, and Fig. 6 shows the variation of SNR between the original and reconstructed signals from day 1 across different CRs. As observed, the Wang matrix consistently outperformed the other two measurement matrices, providing higher SNR values across all CRs.

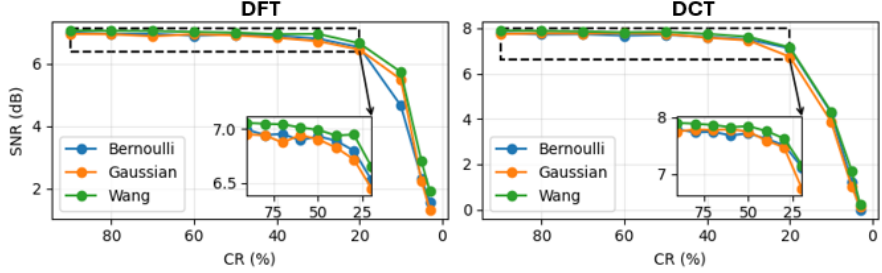

**Fig. 6 SNR vs. CR for a signal from day 1 using OMP in DFT (left) and DCT (right) domains with three measurement matrices (Bernoulli, Gaussian, Wang)**

Fig. 7 illustrates the comparison between the original and reconstructed vibration signals from days 1 and 12, obtained using the DFT basis, Wang matrix, and OMP algorithm at a CR of 10%. The reconstructed waveforms closely follow the originals, preserving key temporal and spectral features. Notably, the reconstructed spectrum exhibits a thresholding-like behavior, which results in the suppression of minor noise components. This behavior can be interpreted as the inherent denoising effect of the CS process. The average SNR across all reconstructed signals is 6.34 dB $\pm$ 0.29 dB, indicating satisfactory reconstruction performance.

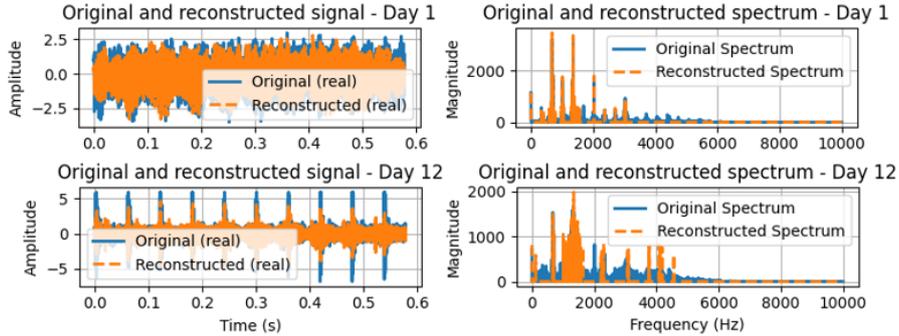

**Fig. 7 Comparison of original and reconstructed signals and spectra (Day 1 and Day 12)**

Furthermore, the kurtosis of each daily signal was computed for both the original and reconstructed datasets. As shown in Fig. 8, the reconstructed signals closely follow the same trend as the original set, demonstrating strong consistency in statistical behavior. Notably, both datasets exhibit a significant rise in kurtosis during the last two days, clearly indicating the onset of a fault. This confirms that kurtosis remains a reliable indicator of signal anomalies even after CS reconstruction using the same parameters (DFT, Wang matrix, OMP, CR = 10%).

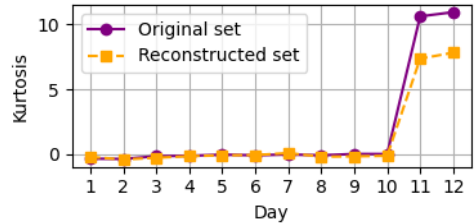

**Fig. 8 Kurtosis comparison between original and reconstructed signals across 12 days**

## 4 Conclusion

This study presented a comprehensive evaluation of compressive sensing (CS) techniques for mechanical vibration signal acquisition. Experiments conducted on the HUMS2023 and CETIM gearbox datasets showed that vibration signals are highly sparse in the DCT and DFT domains, enabling efficient CS-based acquisition. The Wang measurement matrix, introduced for the first time in this context, consistently outperformed conventional Gaussian and Bernoulli matrices by delivering higher SNR, especially at high CRs across both datasets. In addition to its superior reconstruction accuracy, the matrix's hardware-friendly structure enables efficient real-time implementation, making it well-suited for embedded monitoring systems.

## Acknowledgements

Imen Tounsi gratefully acknowledges the European Commission for its support of the Marie Sklodowska Curie program through the Horizon Europe DN PATRON project (GA 101120172)